\documentclass[conference]{IEEEtran}
\IEEEoverridecommandlockouts

\usepackage{cite}
\usepackage{amsmath,amssymb,amsfonts}
\usepackage{algorithmic}
\usepackage{graphicx}
\usepackage{textcomp}
\usepackage{xcolor}
\usepackage[utf8]{inputenc}
\usepackage[T1]{fontenc}
\usepackage{makecell}

\def\BibTeX{{\rm B\kern-.05em{\sc i\kern-.025em b}\kern-.08em
    T\kern-.1667em\lower.7ex\hbox{E}\kern-.125emX}}
\begin{document}

\title{Enabling Adversarial Robustness in AI Models through Kubeflow MLOps\\

}



\author{
\IEEEauthorblockN{
Stavros Bouras$^{1}$,
Ioannis Korontanis$^{2}$,
Antonios Makris$^{1}$,
Konstantinos Tserpes$^{1}$
}
\IEEEauthorblockA{\textit{$^{1}$School of Electrical and Computer Engineering, National Technical University of Athens, Greece}}
\IEEEauthorblockA{\textit{$^{2}$Department of Informatics and Telematics, Harokopio University of Athens, Greece}}
\IEEEauthorblockA{\{stavros\_bouras, antoniosmakris, tserpes\}@mail.ntua.gr, gkorod@hua.gr}
}

\maketitle

\begin{abstract}
AI models are increasingly deployed in cloud-native environments to support scalable and automated services. However, while platforms such as Kubernetes provide strong infrastructure orchestration, security mechanisms specifically designed to protect deployed AI models remain limited. This paper presents security measures for AI models deployed in Kubernetes clusters. The proposed architecture integrates Kubeflow-based MLOps to automatically detect adversarial attacks during the inference phase and trigger defense mechanisms that preserve the model’s accuracy and reliability. Specifically, a Fast Gradient Sign Method (FGSM) attack is applied at inference time, and a Projected Gradient Descent (PGD)-based adversarial training defense is automatically deployed when a degradation in accuracy is detected. The experimental results indicate that the deployed defense robustifies the model, significantly recovering accuracy relative to the degradation caused by the attack.
\end{abstract}

\begin{IEEEkeywords}
Adversarial Robustness, Adversarial Machine Learning, Adversarial Attacks, Security, MLOps 
\end{IEEEkeywords}

\section{Introduction}


The use of containerization and orchestration technologies offers numerous benefits for deploying cloud-native applications. Kubernetes is currently the leading container orchestration platform to deploy and scale multi-component applications, deploying them as a set of pods. In Kubernetes, the main built-in security mechanisms are RBAC (Role-Based Access Control) and network policies. RBAC restricts what actions users and applications within the cluster can perform, enforcing the principle of least privilege and helping prevent unauthorized access or privilege escalation. Network policies control the flow of network traffic between pods and namespaces, limiting communication paths and reducing the attack surface. These mechanisms are defined and configured by the cluster administrator, who establishes the appropriate roles, bindings, and network rules to enforce security policies.

Both \cite{10959185,Kampa_2024} emphasize that proper configuration of the underlying infrastructure in Kubernetes is essential to prevent security breaches. Their studies highlight that common security issues may arise from unverified container images, vulnerabilities in the Kubelet API, poor cluster configuration, insufficiently secure network policies, use of default Kubernetes settings, improperly configured RBAC (Role-Based Access Control) policies, and insecure storage of Kubernetes Secrets. These misconfigurations can lead to multiple types of attacks, including container compromise, privilege escalation, unauthorized access to cluster resources, lateral movement across nodes, data exfiltration, and denial-of-service (DoS) attacks, posing significant risks to both the integrity and availability of the system.


Kubernetes itself is not designed to defend against emerging attack vectors targeting Machine Learning (ML) deployed models, and the platform alone lacks mechanisms to fully mitigate such threats. Instead as it was previously mentioned, the security level of Kubernetes is primarily focused on infrastructure-level concerns such as network and node security. 
Prior work \cite{makris2025coevolution} highlighted the importance of deploying ML models as components of interconnected systems rather than in isolation.
A representative example of such an interconnected system is the integration of MLOps tools with Kubernetes to enable automated pipelines that manage the complete lifecycle of ML models and deploy them to production-ready environments for inference. In such environments, models may be susceptible to adversarial attacks and exhibit vulnerabilities during both the training and inference phases, making the need for defensive methods that enhance their robustness evident. Based on this motivation, our paper presents a method for incorporating adversarial robustness into AI models through the MLOps tool Kubeflow\footnote{https://www.kubeflow.org/}. Kubeflow is an open-source platform used to build, train, and deploy ML models on Kubernetes. Our approach demonstrates how Kubeflow can be used to automate adversarial attack tracking during the inference phase and apply automated defense pipelines.

The remainder of this paper is organized as follows. Section \ref{sec:related_work} presents a literature review on Kubernetes attacks and defenses. Section \ref{sec:background} provides background on both Kubeflow and the adversarial techniques used in the proposed approach. Section \ref{sec:methodology} introduces the proposed architecture and thoroughly examines how adversarial techniques are integrated into Kubeflow. This section also presents the results of our experiments and the experimental. Finally, Section \ref{sec:conclusions} discusses the impact of the proposed Kubeflow MLOps architecture on securing models in a Kubernetes environment and explores potential future improvements.

\section{Related Work}
\label{sec:related_work}

A review of the existing literature indicates that most research on Kubernetes security primarily focuses on protecting the cluster's physical or virtual machines, stored data, and network quality of service. To the best of our knowledge, there is currently limited research on defense techniques specifically designed to protect AI models deployed within Kubernetes. Therefore, existing research primarily focuses on defending against node-level attacks, network-level attacks, and insider threats.

Based on \cite{gajbhiye2024managing}, the kind of above mentioned threats, should be defended in different phases. The authors propose a conceptual framework that is able to perform vulnerability assessment, identifying weaknesses through container image scanning, Kubernetes analysis, and dependency checks. Next is risk analysis, which prioritizes vulnerabilities using Common Vulnerability Scoring System (CVSS) based on their severity and impact. The third phase focuses on mitigation, including image hardening, runtime monitoring, and strengthened Kubernetes configurations (e.g., RBAC and network policies). This is followed by continuous monitoring for real-time threat detection, updates, and compliance. Finally, incident response defines procedures for handling and reviewing security incidents to improve the overall security of Kubernetes.

Center for Internet Security (CIS) benchmark, a widely recognized set of security guidelines for strengthening Kubernetes nodes, is the standard approach for defending against node-level attacks. Rodriguez et al. A framework combining AI-driven certificate management with CIS-benchmarked configurations is proposed in \cite{rodriguezcontinuous}, achieving 98.7\% certificate rotation accuracy, a 92.4\% reduction in configuration drift, and 99.96\% cluster availability. This approach is extended in \cite{thornetowards} through automated CIS enforcement, AI-driven secrets management, and ML-based autonomous recovery, forming a closed-loop system that ensures compliance and cluster resilience. Similarly, \cite{vance1unified} present an AI-driven framework integrating continuous CIS enforcement, predictive certificate and secrets management, and ML-based cluster healing, with prototypes demonstrating a 92\% reduction in configuration drift, 99.9\% prevention of certificate expiry incidents, and an 85\% decrease in Mean Time to Recovery (MTTR).

Network-level attacks are among the most prevalent threats in Kubernetes environments, and numerous studies have proposed defenses against them. In \cite{10844916}, GrassHopper, a dynamic cross-layer enforcement system, is presented, aligning VM-level Security Group rules with container-level network policies and scheduling decisions. By automatically updating firewall configurations, GrassHopper ensures consistency across layers. Experimental evaluation on a Kubernetes cluster deployed on OpenStack demonstrates that GrassHopper reduces the network attack surface between VMs by 78-99\% without incurring noticeable latency or throughput overhead, effectively mitigating lateral movement while preserving application performance.

There are tools designed to protect Kubernetes from multiple classes of attacks. For example, in \cite{aly2025real}, a framework to enhance security is proposed. The framework integrates KServe\footnote{https://kserve.github.io/website/} to deploy scalable ML-based threat detection, CICFlowMeter\footnote{https://github.com/ahlashkari/CICFlowMeter} for network traffic feature extraction, and KubeDeceive \cite{article} for dynamically deploying decoys when suspicious activity is detected. These components operate within a MAPE-K loop, enabling continuous monitoring and adaptation. In this approach, ML serves as the core defense mechanism, allowing the system to detect and classify various attack types, including reconnaissance, privilege escalation, and denial-of-service attacks. Experimental results show detection accuracy of up to 91\% and decoy success rates reaching 93\%, demonstrating the framework's effectiveness in mitigating threats and improving overall system resilience.

A comparable multilayered defense tool is Kubehound, introduced in \cite{fi15070228}, which is designed to automatically identify security weaknesses in microservice applications deployed on Kubernetes. The tool combines static analysis, examining source code, OpenAPI specifications, and configuration files, with dynamic analysis that scans live clusters, enabling comprehensive detection of security issues. These issues include insufficient access control, exposed services and open ports, overprivileged pods, misconfigured network policies, hardcoded secrets or unencrypted sensitive data, and authentication flaws including token mismanagement. By systematically detecting these vulnerabilities, Kubehound enables organizations to proactively  mitigate microservice security risks before they can be exploited.

All the aforementioned works focus on protecting the Kubernetes infrastructure at the node and network levels, but they do not address the security of the AI models themselves. This paper addresses this gap by incorporating the adversarial robustness of the deployed AI models as an automated capability within the Kubeflow environment.

\section{Background}
\label{sec:background}
\subsection{Kubeflow and MLOps Pipelines}

Kubeflow enables automated, end-to-end orchestration of ML workflows on Kubernetes, supporting distributed model training and deployment as APIs or batch jobs. Its Pipelines Domain Specific Language (DSL) allows developers to define workflows in Python, specifying Docker images through component decorators and defining the execution sequence with a pipeline decorator, including any external dependencies beyond the application code. By leveraging Kubernetes' orchestration capabilities, Kubeflow Pipelines ensures automation, scalability and reproducibility, while enabling seamless integration of data processing, model evaluation, deployment, and monitoring. The trained model is produced as the final output of the workflow.

\subsection{Adversarial Machine Learning}

ML models are widely deployed in AI applications, yet they remain vulnerable to adversarial attacks. Adversarial Machine Learning (AML) \cite{huang2011adversarial} is a subfield of ML that focuses on identifying vulnerabilities in model behavior and studying both the design of adversarial attacks that exploit these weaknesses and the development of defenses to enhance model robustness and security. The ML life-cycle consists of two major phases, the training phase and the inference phase. Unlike traditional ML, which assumes trusted training and inference, AML considers scenarios in which an adversary deliberately manipulates data or inputs to compromise model behavior. Both phases may be targeted by adversarial attacks, with defenses deployed at each stage to improve model robustness.

\subsubsection{Adversarial Attacks}

A thorough description of adversarial attacks requires the clear definition of three key dimensions \cite{zhou2022adversarial}: 
\begin{itemize}
  \item Adversary's Goal: Depending on the objective of the adversary, the attacks can be broadly grouped into \textit{poisoning}, \textit{evasion} and \textit{privacy/inference} attacks. These attacks occur at different phases of the AI life-cycle, with poisoning attacks targeting the training phase, while evasion and privacy/inference attacks occur during the inference-phase. In \textit{poisoning} attacks, adversaries have access and can modify the training data in order to influence the behavior of the model during inference \cite{biggio2012poisoning,jagielski2018manipulating}. In \textit{evasion} attacks, attackers target trained ML models without altering their training data or parameters, crafting inputs to mislead the model \cite{szegedy2013intriguing,goodfellow2014explaining}. Finally, in \textit{privacy/inference} attacks, adversaries aim to extract information about the model, such as features of training data \cite{fredrikson2014privacy}, model parameters \cite{tramer2016stealing, wang2018stealing} and sensitive attributes \cite{shokri2017membership}.
  \item Adversarial Specificity: Depending on whether the attacker aims to induce a specific incorrect model output or simply any incorrect output, adversarial attacks can be further categorized into \textit{targeted} and \textit{untargeted} respectively \cite{zhou2022adversarial}. In a \textit{targeted} attack, the adversary seeks to force the model to produce a specific, pre-defined incorrect output. For example, in a classification setting, this may correspond to misclassifying an input as a chosen target class. In contrast, an \textit{untargeted} attack aims only to cause an incorrect output, without constraining the model to any particular outcome.
  \item Adversary's Knowledge/Capabilities: Based on the information and resources available to the attacker, adversarial attacks are grouped into \textit{white-box}, \textit{gray-box} and \textit{black-box} \cite{zhang2025adversarial}. In the \textit{white-box} setting, the adversary has complete access to the target model along with its architecture, training dataset and parameters. In the \textit{gray-box} setting, the adversary has partial knowledge and access of the victim model. Finally, in the \textit{black-box} setting the adversary has no knowledge regarding the victim model and its training data, architecture and parameters.
\end{itemize}

\subsubsection{Adversarial Defenses}

To mitigate adversarial threats, defense mechanisms have been developed to enhance the robustness of ML models. Similar to adversarial attacks, these defenses can be applied either during the training phase or at inference phase. Training-time defense mechanisms aim to improve the model resilience during training by exposing it to adversarial knowledge in its learning process, whereas inference-time defenses aim to protect the deployed model from adversarial inputs during deployment. A well-established example of a training-time defense is \textit{adversarial training} \cite{goodfellow2014explaining, madry2017towards}, which involves generating adversarial examples during training to improve model robustness. At the inference stage, a commonly used defense is input reconstruction/purification, where potentially adversarial inputs are transformed before being processed by the model \cite{samangouei2018defense, nie2022diffusion}.

\section{Methodology}
\label{sec:methodology}

\subsection{System Architecture}

The main methodology of this paper involves integrating AML into an architecture built on Kubeflow. The proposed setup builds on our previous work, which is based on the training architecture of Korontanis et. al \cite{10.1145/3770501.3771304}. In this design, Kubeflow serves as the core component responsible for orchestrating model training workflows, while also providing a dataset registry that allows users to store and manage datasets used during training. In addition, the setup leverages the Kubernetes NVIDIA plugin and Kyverno to enable temporary pods spawned by Kubeflow to access GPU resources.


\subsection{Normal usage - without Attack}

\begin{figure}[h!]
    \centering
    \includegraphics[width=0.50\textwidth]{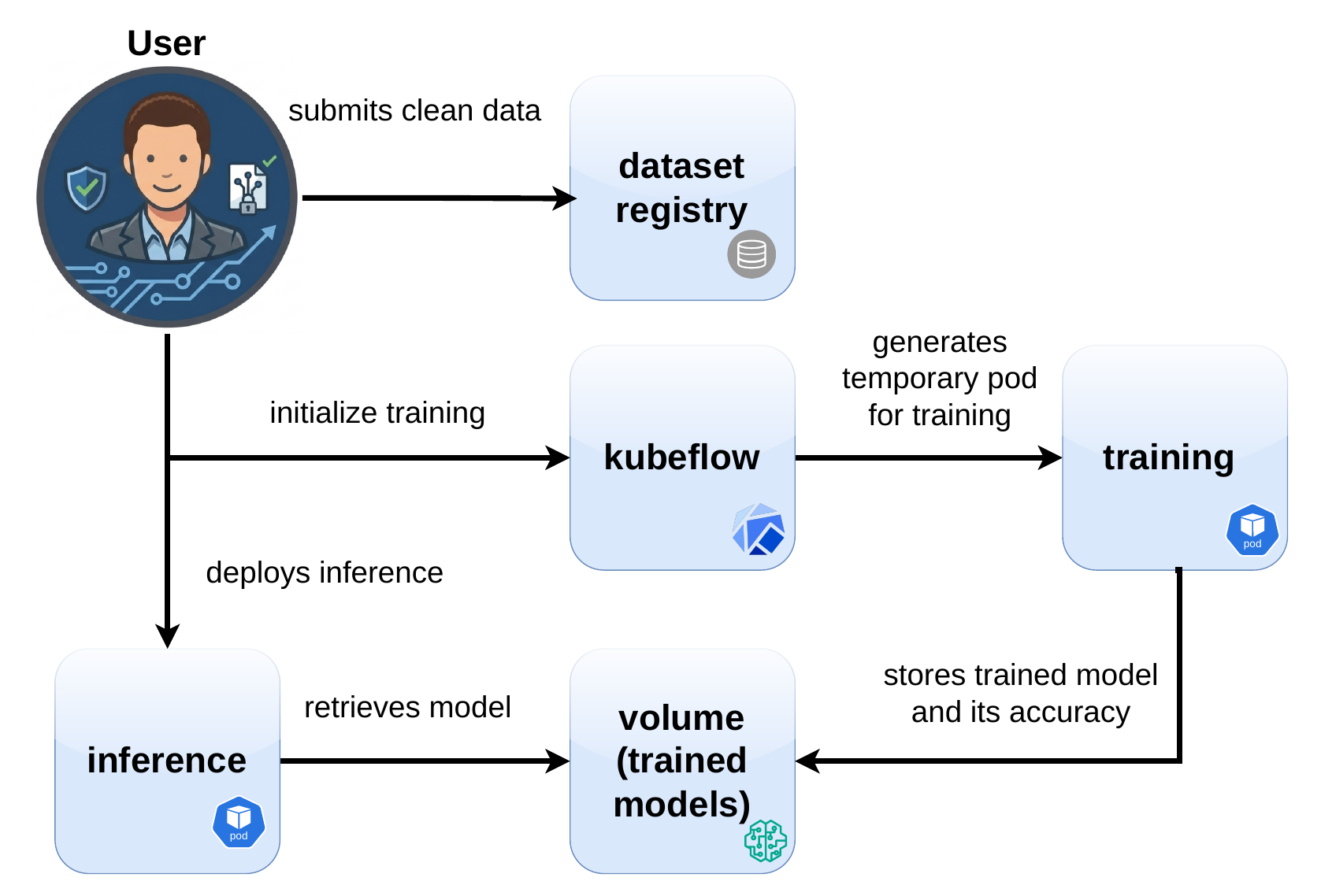}
    \caption{Kubeflow Normal Usage}
    \label{fig:architecture}
\end{figure}

The normal usage of this architecture is illustrated in Figure~\ref{fig:architecture}. Initially, the user uploads a trusted dataset $\mathcal{D}$ to the registry and initiates model training through Kubeflow. This process spawns a temporary pod that trains a CNN classifier $f_A$ on $\mathcal{D}$ by minimizing a loss function $\mathcal{L}$. Upon completion, the trained model $f_A$ and its clean baseline accuracy $\alpha_A$ are stored in a Kubernetes persistent volume, where $\alpha_A$ serves as the system’s integrity baseline. The user then deploys an inference pod, which retrieves the model and initiates the inference process.
        
\subsection{Adversarial scenario - Adversarial Examples}
\begin{figure}[h!]
    \centering
    \includegraphics[width=0.40\textwidth]{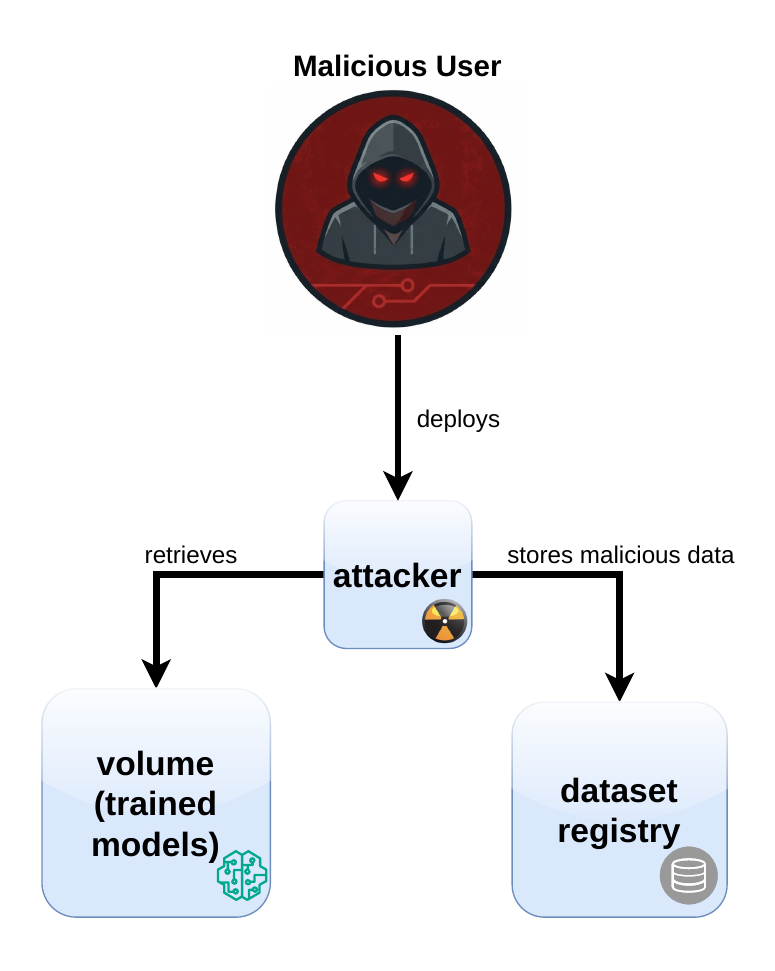}
    \caption{Kubeflow Adversarial Scenario}
    \label{fig:attack}
\end{figure}

The adversarial scenario is illustrated in Figure 2 and represents an insider threat, where a malicious user has the ability to deploy a pod and retrieve the trained model $f_A$  from the volume (Figure \ref{fig:attack}). The adversary exploits $f_A$ in order to generate an adversarial dataset $\mathcal{D}_{\text{adv}}$, consisting of adversarial examples that will be uploaded to the dataset registry and will deteriorate the accuracy of the deployed model. For the adversarial examples to be generated, the Fast Gradient Sign Method (FGSM) \cite{goodfellow2014explaining} adversarial attack is applied to the clean dataset $\mathcal{D}$. As a result, FGSM crafts adversarial examples by perturbing the samples of $\mathcal{D}$ in the direction of the loss gradient, as shown in the following equation:

\begin{equation}
    x_{\text{adv}} = x + \varepsilon \cdot \text{sign}\left(\nabla_x 
    \mathcal{L}(f_A(x;\theta_A), y)\right)
    \label{eq:fgsm}
\end{equation}

\noindent where $x$ is a clean input sample, $y$ is its ground truth label, 
$\theta_A$ are the parameters of the baseline model, and $\varepsilon$ 
is the magnitude of the perturbation.

\subsection{Defense scenario - Adversarial Training}

The defense mechanism implemented in this system architecture relies on adversarial training, as presented in \cite{madry2017towards}. This work frames adversarial training as a saddle point (min-max) optimization problem during which an inner maximization problem identifies perturbations that maximize a loss for a given input and an outer minimization problem finds parameters $\theta$ that minimize the loss with respect to $\theta$. Formally this is described in the following equation:
\begin{equation}
\min_{\theta} \ \mathbb{E}_{(x,y)\sim D} \left[ \max_{\delta \in S} L(\theta, x + \delta, y) \right]
\end{equation}

\noindent where $\theta$ refers to the model parameters, $D$ to the data distribution over the inputs $x$ and their corresponding labels $y$, $L$ the loss function, and $S$ defines the set of allowed perturbations from which $\delta$ is optimized. The inner optimization problem is solved using Projected Gradient Descent (PGD) as presented below: 
\begin{equation}
x^{t+1} = \Pi_{x+S}\left(x^t + \alpha \,\mathrm{sgn}\!\left(\nabla_x L(\theta, x^t, y)\right)\right)
\end{equation}

\noindent where $x^{t+1}$ is the updated adversarial sample at iteration $t+1$ given $x^t$,  $\nabla_x L(\theta, x^t, y)$ is the gradient of the loss with respect to the input, $\alpha$ is the step size and $\Pi_{x+S}$ is the projection onto the allowed perturbation set $S$. The allowed perturbation set $S$ is the set of perturbations $\delta \in \mathbb{R}^d$ such that $\|\delta\|_\infty \le \epsilon$. The parameter $\epsilon$ defines the perturbation budget, controlling the radius of the $\ell_\infty$-ball around each input sample and determining the maximum allowable perturbation of each sample during adversarial training.

\begin{figure}[!h]
    \centering
    \includegraphics[width=0.45\textwidth]{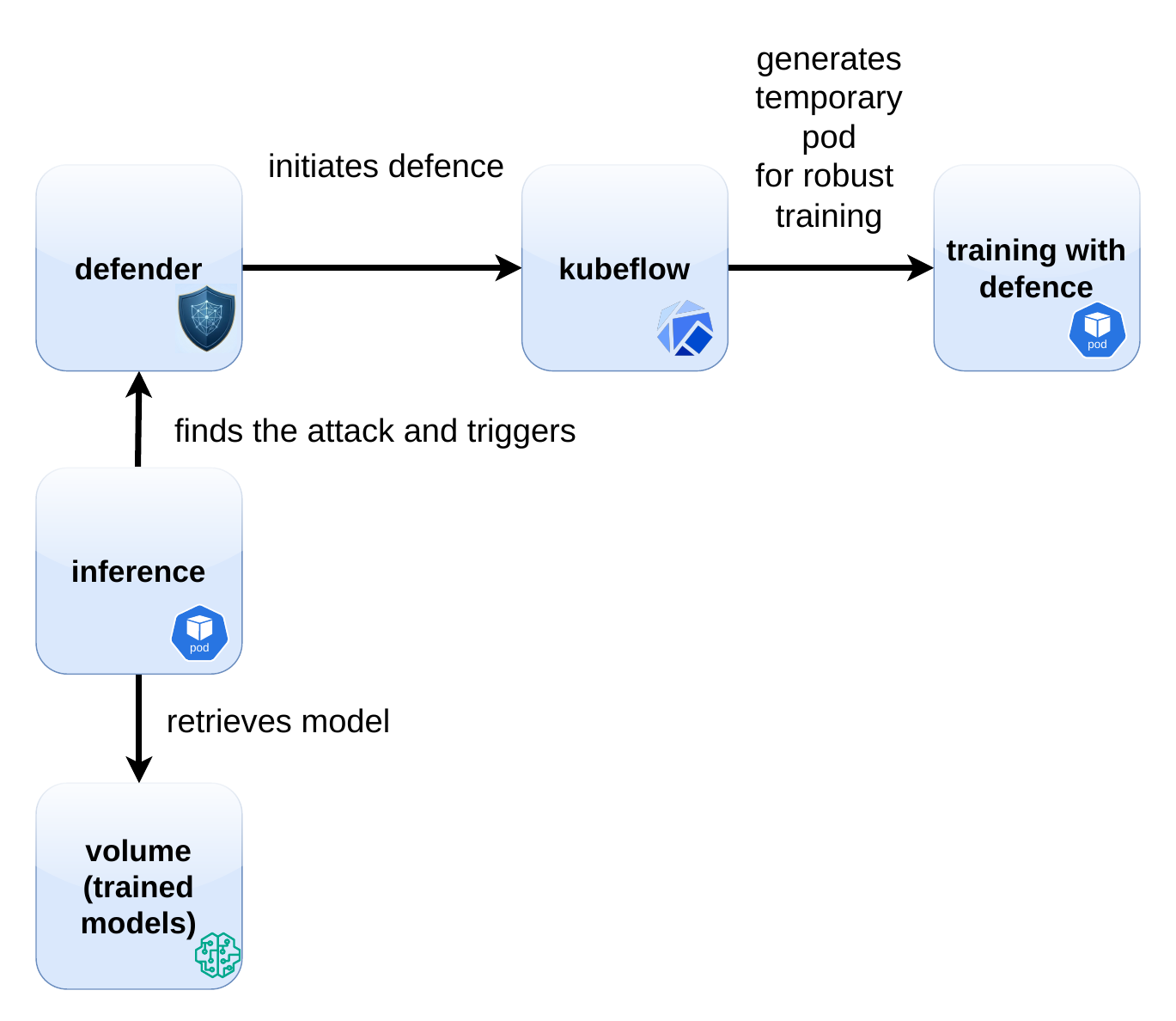}
    \caption{Kubeflow Defence Scenario}
    \label{fig:defence}
\end{figure}

The defense scenario is shown in Figure~\ref{fig:defence} and is  triggered during the inference phase. The inference pod monitors the accuracy of $f_A$ on incoming data and compares it against the stored baseline accuracy $\alpha_A$. If a deterioration of more than 5\% in the reported accuracy is reported, the inference pod triggers the defender component, which instructs Kubeflow to initiate an adversarial training pipeline. A temporary pod is being spawned by Kubeflow that loads $f_A$ from the volume and performs the defense.




\section{Implementation and Results}

The adversarial robustness is evaluated under varying attack and defense scenarios. The baseline classifier $f_A$ is a Convolutional Neural Network (CNN) \cite{li2021survey} trained on the MNIST handwritten digit classification dataset. In addition, the hardened model $f_{A'}$ was initialized from the weights of $f_A$ and fine-tuned using adversarial training using PGD-generated adversarial examples. Two scenarios were conducted for evaluation, a white-box attack where FGSM was applied against $f_{A'}$, and a transfer attack where $f_{A'}$ was evaluated with adversarial examples generated from $f_A$, simulating a setting where the adversary does not have knowledge of the deployed hardened model and defense. 

The experimental setup consisted of both fixed and variable parameters. The fixed parameters that were kept constant across all experiments are shown in Table~\ref{tab:fixed_params}. 

\begin{table}[h]
\centering
\begin{tabular}{l c l}
\textbf{Parameter} & \textbf{Value} & \textbf{Description} \\
\hline
\noalign{\vskip 2pt}
Epochs ($f_A$) & 10 & Baseline training \\
Epochs ($f_{A'}$) & 20 & Adversarial retraining \\
PGD steps ($k$) & 20 & Iterations per batch \\
Step size ($\alpha$) & 0.01 & PGD step size \\
Learning rate ($f_A$) & 0.001 & Baseline training \\
Learning rate ($f_{A'}$) & 0.0001 & Fine-tuning \\
\hline
\end{tabular}
\vspace{0.2cm}
\caption{Fixed parameters across all experiments.}
\label{tab:fixed_params}
\end{table}

Having the aforementioned variables fixed, the variables that varied across the different experiments was the magnitude of the perturbation $\varepsilon$ in the FGSM attack and the perturbation budget $\epsilon$ of the PGD-adversarial training, as show in Table~\ref{tab:variable_params}.

\begin{table}[h]
\centering
\begin{tabular}{l c}
\textbf{Parameter} & \textbf{Values tested} \\
\hline
\noalign{\vskip 2pt}
Attack: FGSM perturbation magnitude $\varepsilon$ & $\{0.15,\ 0.20,\ 0.25\}$ \\
Defense: PGD perturbation budget $\epsilon$ & $\{0.15,\ 0.20,\ 0.25,\ 0.30\}$ \\
\hline
\end{tabular}
\vspace{0.2cm}
\caption{Variable parameters explored across experiments.}
\label{tab:variable_params}
\end{table}

\begin{figure}[t!]
\centering
\includegraphics[width=\columnwidth]{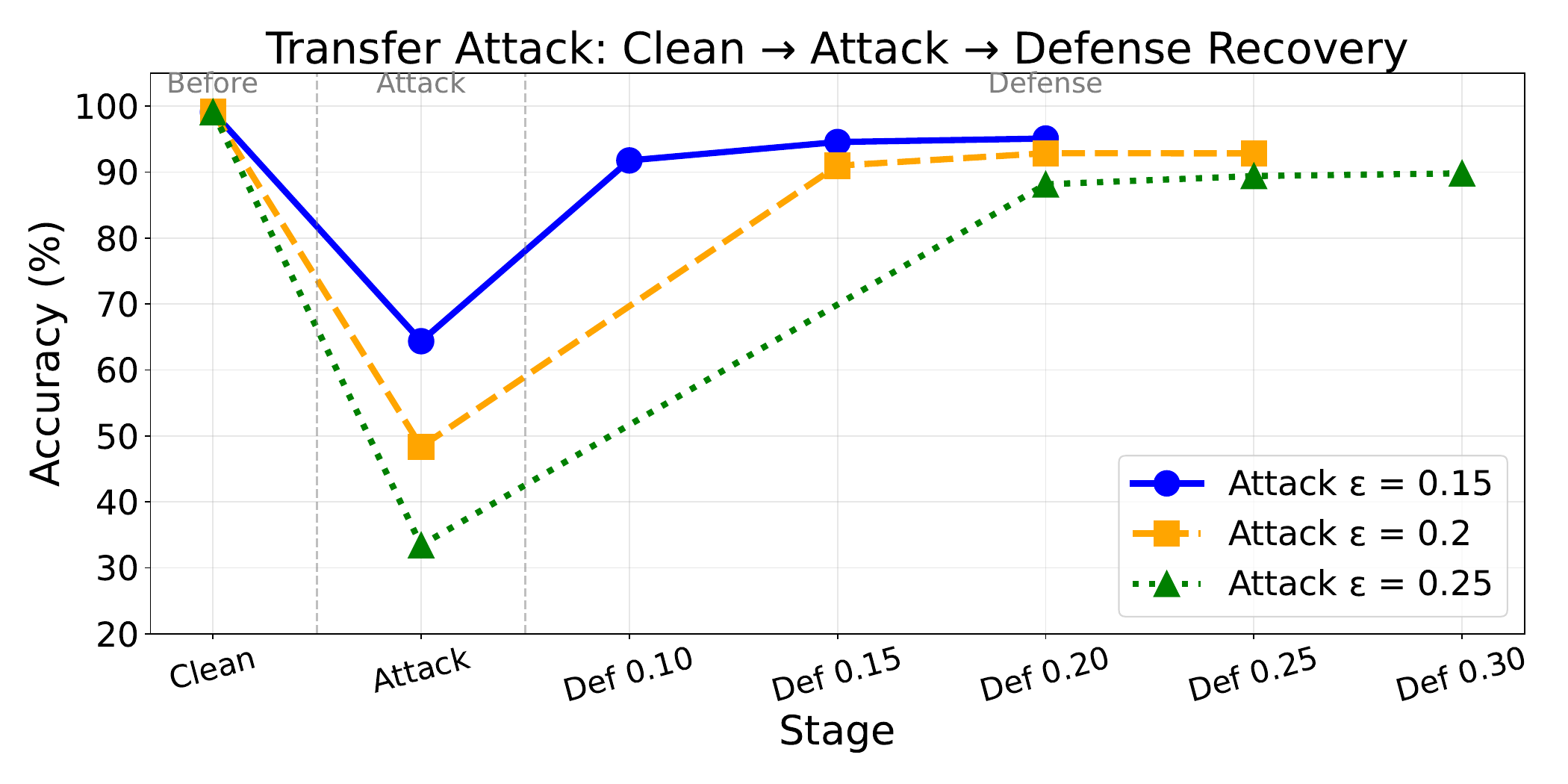}
\caption{Accuracy evolution from clean inputs to adversarial attack and subsequent defense recovery in the transfer setting (FGSM from $A$ on $A'$).}
\label{fig:transfer}
\end{figure}

\begin{figure}[t!]
\centering
\includegraphics[width=\columnwidth]{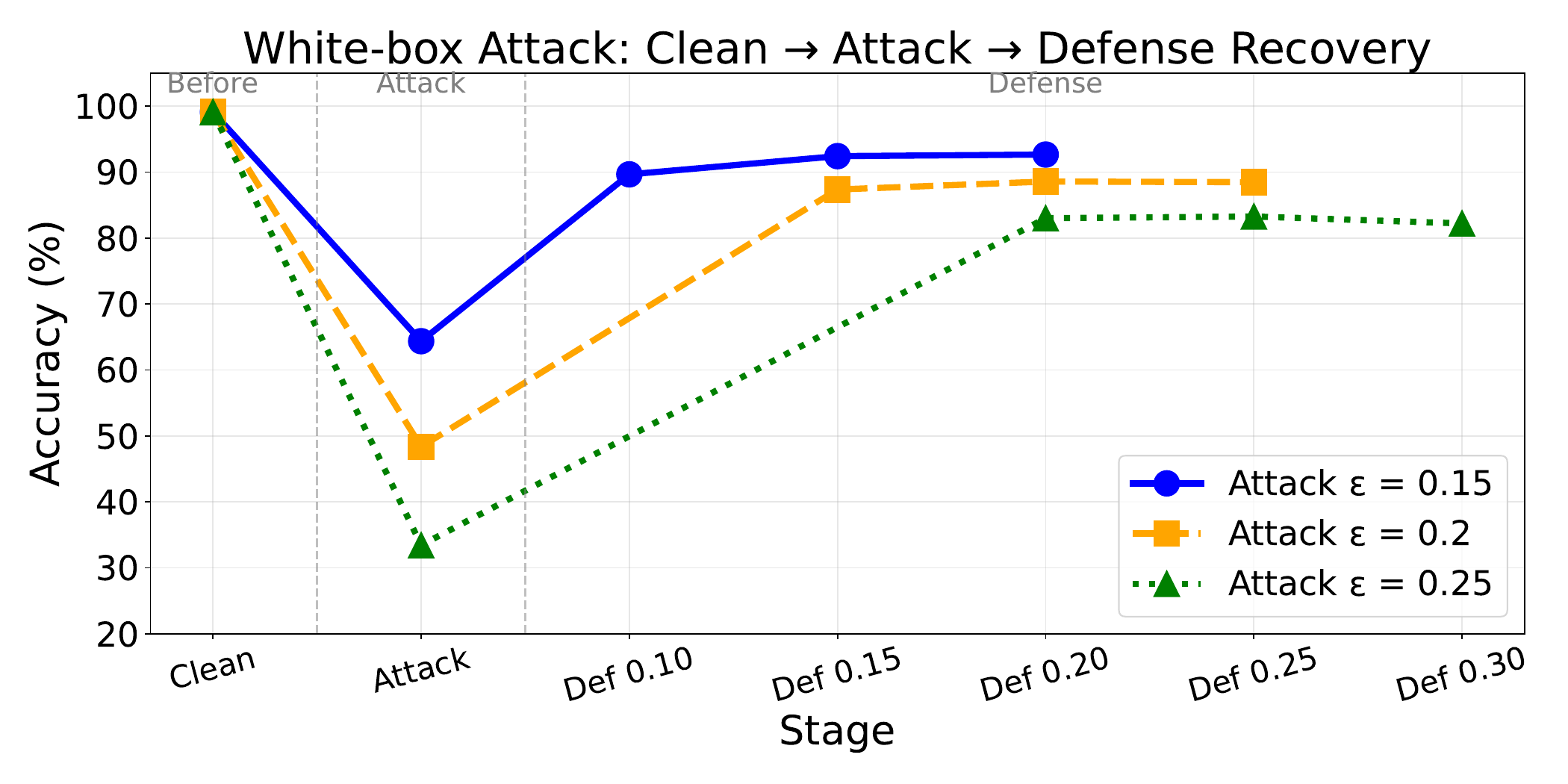}
\caption{Accuracy evolution from clean inputs to adversarial attack and subsequent defense recovery in the white-box setting (FGSM from $A'$ on $A'$).}
\label{fig:whitebox}
\end{figure}

The conducted experiments report white-box (FGSM from $A'$ on $A'$) and transfer (FGSM from $A$ on $A'$) classification accuracies across different settings that simulate varying attack and defense strengths, characterized by the FGSM perturbation magnitude $\epsilon$ and the PGD perturbation budget $\epsilon$. Specifically, three attacking scenarios were tested described with three different FGSM perturbation magnitudes. For each attacking scenario, the PGD perturbation budget used in the corresponding defense was varied around the attack perturbation magnitude, allowing the evaluation of defense configurations with smaller, equal, or larger perturbation budgets relative to the attack. In addition, the baseline accuracy (FGSM on $A$) is reported to show the degradation in performance under adversarial attack, while the clean accuracy of $A'$ is also reported, indicating that the robustified model maintains high performance on unperturbed inputs. All the results are presented in Table~\ref{tab:results}, while Figures~\ref{fig:transfer} and~\ref{fig:whitebox} illustrate the accuracy evolution from clean inputs through adversarial attack to defense recovery in the transfer and white-box settings, respectively.

\begin{table}[t]
\centering
\begin{tabular}{c c c c c c}
\makecell{\textbf{Attack} \\ $\varepsilon$} &
\makecell{\textbf{Defense} \\ $\epsilon$} &
\makecell{\textbf{FGSM} \\ on $A$} &
\makecell{\textbf{FGSM} \\ from $A$ on $A'$} &
\makecell{\textbf{FGSM} \\ from $A'$ on $A'$} &
\makecell{\textbf{Clean} \\ $A'$} \\
\hline
\noalign{\vskip 2pt}
0.15 & 0.10 & 64.36\% & 91.77\% & 89.66\% & 99.37\% \\
0.15 & 0.15 & 64.36\% & 94.56\% & 92.42\% & 99.18\% \\
0.15 & 0.20 & 64.36\% & \textbf{95.09\%} & \textbf{92.66\%} & 98.96\% \\\\
0.20 & 0.15 & 48.36\% & 90.98\% & 87.36\% & 99.22\% \\
0.20 & 0.20 & 48.36\% & \textbf{92.86\%} & \textbf{88.57\%} & 98.96\% \\
0.20 & 0.25 & 48.36\% & 92.84\% & 88.47\% & 98.74\% \\\\
0.25 & 0.20 & 33.43\% & 88.15\% & 83.01\% & 99.00\% \\
0.25 & 0.25 & 33.43\% & 89.40\% & \textbf{83.27\%} & 98.83\% \\
0.25 & 0.30 & 33.43\% & \textbf{89.81\%} & 82.22\% & 98.55\% \\
\hline
\end{tabular}
\vspace{0.2cm}
\caption{Classification accuracy (\%) across different attack and defense strengths.}
\label{tab:results}
\end{table}

Our experiments demonstrate that when the defense budget is equal or slightly over the attack magnitude, the model achieves the highest robustness against both transfer and white-box attacks. When the defense budget is lower than the attack magnitude, the deployed defense performs worse, as it was not trained against perturbations as large as those encountered during inference. These results show that the defense should be calibrated, in this setting, to match the corresponding attack strength in order to achieve optimal performance.

\section{Conclusions \& Future work}
\label{sec:conclusions}
This paper presents an approach for integrating adversarial robustness capabilities into AI models deployed within a Kubeflow-based MLOps environment. The proposed architecture is based on an existing Kubeflow-based training setup and extends its capabilities in order to automatically detect and respond to adversarial threats. By integrating adversarial attack and defensive capabilities into the Kubeflow pipeline, the proposed architecture automatically detects and responds to adversarial threats, which in this case is an insider attacker that has white-box access and is able to craft perturbations based on the deployed model. The experimental evaluation,  using FGSM attacks and PGD-based adversarial training, demonstrates that the deployed defense robustifies the model, recovering accuracy significantly relative to the degradation caused by the attack under both white-box and transfer settings. The results further show that the defense achieves the highest robustness when its perturbation budget is calibrated to match or slightly exceed the attack magnitude, highlighting the importance of aligning defense strength with the anticipated threat.

As future work, the proposed architecture will be extended to include additional adversarial attacks and defenses targeting both the training and inference phases, securing and evaluating the robustness of AI models across their entire life-cycle. Additionally, integrating node-level and network-level security mechanisms would provide a more holistic defense strategy, ensuring protection of both the underlying Kubernetes infrastructure and the deployed AI models.

\section*{Acknowledgment}

This paper has received funding from the European Union’s Horizon Europe research and innovation actions under grant agreement No 101168560 (CoEvolution). Views and opinions expressed are however those of the author(s) only and do not necessarily reflect those of the European Union or the Commission. Neither the European Union nor the granting authority can be held responsible for them.

\bibliographystyle{IEEEtran}
\bibliography{ref}

\end{document}